\documentclass[floatfix,showpacs,amsmath,amsfonts,amssymb,aps,twocolumn,
superscriptaddress,pra,10pt]{revtex4-1}
\usepackage[pdftex]{graphicx}

\newcommand{\dd}{\mathrm{d}}
\newcommand{\vect}[1]{\mathbf{#1}}
\newcommand{\kf}[0]{k_\mathrm{F}}
\newcommand{\Ef}[0]{E_\mathrm{F}}
\newcommand{\rc}[0]{r_\mathrm{c}}
\newcommand{\expe}[0]{\mathrm{e}}
\newcommand{\Ec}[0]{E_\mathrm{c}}
\newcommand{\kc}[0]{k_\mathrm{c}}
\newcommand{\figref}[1]{Fig.~\ref{#1}}

\newcommand{\reference}[1]{Ref.~\cite{#1}}

\newcommand{\complexi}{\mathrm{i}}

\newcommand{\neweqnline}{\nonumber\\}
\newcommand{\secref}[1]{Section~\ref{#1}}
\newcommand{\eqnref}[1]{Equation~(\ref{#1})}

\newcommand{\scriptA}[0]{\mathcal{A}}
\newcommand{\scriptB}[0]{\mathcal{B}}

\newcommand{\Vcont}[0]{V^\mathrm{cont}}
\newcommand{\re}[0]{R_\mathrm{e}}
\newcommand{\Eb}[0]{E_\mathrm{b}}
\newcommand{\bigO}[0]{\mathcal{O}\!}

\begin{document}

\title{Pseudopotential for the 2D contact interaction}
\author{T.M.~Whitehead}
\affiliation{Cavendish~Laboratory, J.J.~Thomson~Avenue, Cambridge, CB3~0HE, 
United Kingdom}
\author{L.M.~Schonenberg}
\affiliation{Cavendish~Laboratory, J.J.~Thomson~Avenue, Cambridge, CB3~0HE, 
United Kingdom}
\author{N.~Kongsuwan}
\affiliation{Blackett~Laboratory, Prince Consort Road, London, SW7~2AZ, United 
Kingdom}
\author{R.J.~Needs}
\affiliation{Cavendish~Laboratory, J.J.~Thomson~Avenue, Cambridge, CB3~0HE, 
United Kingdom}
\author{G.J.~Conduit}
\affiliation{Cavendish~Laboratory, J.J.~Thomson~Avenue, Cambridge, CB3~0HE, 
United Kingdom}
\date{\today}

\begin{abstract}
We propose a smooth pseudopotential for the contact interaction acting between 
ultracold atoms confined to two dimensions.  The pseudopotential reproduces the 
scattering properties of the repulsive contact interaction up to $200$ times 
more accurately than a hard disk potential, and in the attractive 
branch gives a $10$-fold improvement in accuracy over the square well 
potential.  Furthermore, the new potential enables 
diffusion Monte Carlo simulations of the ultracold gas to be run $15$ times 
quicker than was previously possible.
\end{abstract}

\maketitle

\section{Introduction} 

Many collective quantum phenomena emerge from reduced dimensionality,
including the quantum Hall effect \cite{Lin09},
high-temperature superconductivity \cite{Lee06}, quantum magnetism 
\cite{Sachdev08}, and topological
insulators \cite{Hasan10}.  Consequently, two-dimensional (2D) systems have 
recently attracted a great deal of attention  
\cite{Conduit10,Drummond11,Ngampruetikorn13,Martiyanov10,Sommer12,Gunter05,
Shi15}.
  2D systems may now be realized, for example, at the interface between two 
solids \cite{Bert11}, or in an ultracold atomic gas in an anisotropic optical 
trap, with one dimension tightly confined relative to the other two 
\cite{Gorlitz01,Fenech16,Boettcher16}.  This coincidence of novel many-body 
phenomena with accurate experimental realizations makes 2D systems attractive 
for numerical investigation.

Ultracold atoms provide a clean model Hamiltonian with a tunable
interaction strength, and their study has delivered new insights into
many-body quantum physics
\cite{Hart15,Anderson95,Regal04,Zwierlein04}. The resonant Feshbach
interaction \cite{Chin10} between ultracold atoms is usually modeled
by a contact potential
\cite{Astrakharchik04,Nascimbene10,Bertaina11,Xing90,Bertaina2013,Casula08,
Li11,Bertaina13b,Zhou11}.  Despite its widespread usage, the contact
interaction causes sampling problems in numerical simulations due to
its infinitesimally short range and divergence at coalescence.  It
also harbors a bound state, complicating the use of ground state
methods for examining repulsive scattering between particles.  These
difficulties are conventionally circumvented by replacing the contact
potential by, for example, a hard disk potential, which we show leads
to inaccurate scattering properties.  This problem has recently been
resolved in three dimensions by Bugnion \emph{et al.} with the
development of a smooth pseudopotential \cite{Bugnion14} that results
in a hundred-fold increase in the accuracy of the scattering
properties.  The smoothness of the new pseudopotential also radically
speeds up numerical calculations
\cite{Lloyd-Williams2015,Whitehead16}. Here we follow that
prescription to develop a pseudopotential that improves the modeling
of 2D quantum gases with a contact interaction.

In \secref{sec:contact} we discuss two particles interacting
via the 2D contact potential.  In \secref{sec:derivation} we derive
the pseudopotential, and in
\secref{sec:harmonic} demonstrate its
accuracy in an inhomogeneous two-body
system.  In \secref{sec:fermi} we examine the pseudopotential's advantages over 
other methods in a homogeneous many-body system, before discussing 
potential future applications of the pseudopotential in 
\secref{Sec:discussion}.

\section{Analytical results}
\label{sec:contact}

In order to develop a pseudopotential for use in many-body simulations, 
it is essential to first understand the behavior of the
two-particle system.  Here we analyze an isolated two-body system of
distinguishable fermions, starting with non-interacting particles and
then adding a short-ranged interaction potential, which not only allows us to 
find solutions for the contact interaction, but also serves as a platform from 
which to propose a pseudopotential.  Atomic units (\mbox{$\hbar=m=1$}) are used
throughout, and anticipating that we will
study many-body systems, we measure energies in units of the Fermi
energy $\Ef$ and lengths in units of the Fermi length $\kf^{-1}$.

\subsection{Short-ranged two-particle interactions}

We consider two equal-mass, distinguishable fermions in a vacuum.  In their 
center-of-mass frame, the Schr\"odinger Equation for particles
interacting via a potential $V(r)$ is given by
\begin{align}
-\nabla^2 \psi(r,\theta) + V(r) \psi(r,\theta) = E \psi(r,\theta) ,
\label{eq:SchrEq}
\end{align}
where $E$ is the energy in the center-of-mass frame.

The analytic solution to \eqnref{eq:SchrEq} for non-interacting
particles ($V(r)=0$) in a vacuum takes the form
\begin{align*}
\psi_\ell (r,\theta) = R_\ell (r) \Theta_\ell(\theta)
\end{align*}
with 
\begin{align*}
\Theta_\ell (\theta)&=\frac{1}{\sqrt{2\pi}} \expe^{\complexi \ell \theta}, \\
R_\ell (r) &= \scriptA(k) J_\ell (k r) + \scriptB(k) Y_\ell (k r),
\end{align*}
where $k=\sqrt{E}$ is the wavevector in the center-of-mass frame,
$\ell$ is angular momentum projected onto the normal to the 2D plane,
and $\scriptA(k)$ and $\scriptB(k)$ are constants set by the boundary 
conditions. $J_\ell (kr)$ and $Y_\ell (kr)$ are Bessel functions of the
first and second kinds, respectively.

If we take the potential $V(r)$ to be short ranged and
cylindrically symmetric, for distinguishable fermions the only effect of the 
potential is in the $\ell=0$ channel.  The wavefunction beyond the interaction 
range, where $V(r)=0$, then takes the same form as in the non-interacting case,
\begin{align}
\psi_0(r)\propto \scriptA(k) J_0(k r) + \scriptB(k) Y_0 (kr).
\label{eq:nonintsolution}
\end{align}
There are two branches of solutions, scattering states for
$E>0$ and bound states for $E<0$.

\subsubsection{Scattering states ($E>0$)}
For two-particle scattering with positive $E$, the non-interacting wavefunction 
given by \eqnref{eq:nonintsolution} with $k=\sqrt{E}$ can be written at large 
separations in the oscillatory form
\begin{align}
  \psi_\mathrm{s}(r) \propto \frac{\sin\left(kr +
      \frac{\pi}{4}+\delta(k)\right)}{\sqrt{kr}},
\label{eq:scatteringwfn}
\end{align}
where the scattering phase shift $\delta(k)$, given by
\begin{align}
\mathrm{cot}\, \delta=-\scriptA(k)/\scriptB(k),
\label{eq:phaseshift}
\end{align}
describes the large radius behavior of the wavefunction and captures the full 
impact of the scattering interaction.

\subsubsection{Bound states ($E<0$)}
Two particles with $E<0$ are in a bound state in which they
remain in close proximity if no external force is applied.  The bound
state wavefunction has the form
\begin{align*}
  \psi_\mathrm{b}(r)\propto \scriptA(\kappa)J_0(\complexi \kappa r) +
  \scriptB(\kappa) Y_0(\complexi \kappa r),
\end{align*}
where $\kappa=\sqrt{-E}$.  For the wavefunction to be normalizable
$\scriptA$ and $\scriptB$ must satisfy
$\scriptB(\kappa)/\scriptA(\kappa)=\complexi$, and therefore the
wavefunction
\begin{align}
  \psi_\mathrm{b}(r)&\propto J_0(\complexi \kappa r) + \complexi Y_0
  (\complexi \kappa r) \neweqnline &\propto H_0^{(1)}(\complexi \kappa
  r),
\label{eq:boundwfn}
\end{align}
where $H_0^{(1)}(x)=J_0(x)+\complexi Y_0(x)$ is the Hankel function of
the first kind.  Note that $H_0^{(1)}(\complexi \kappa r)\to
\expe^{-\kappa r}/\sqrt{\kappa r}$ as $\kappa r \to \infty$, with 
the expected exponential decay of a bound state.

\subsection{2D contact interaction}

We now apply these results for short-ranged 2D interactions to the 2D
contact interaction $\Vcont(r)$.  In a fermionic system this
zero-ranged interaction acts only between distinguishable
particles, with the interaction strength described by a scattering
length $a$.  We can capture the full effect of the
interaction by imposing the boundary condition \cite{Liu10,Farrell10}
\begin{align}
\left( r \frac{\dd}{\dd r} - \frac{1}{\mathrm{ln}(r/a)} \right)\psi(r)=0
\label{eq:boundary}
\end{align}
at $r=0$ and then at $r>0$ use the non-interacting solution 
\eqnref{eq:nonintsolution}, which gives
\begin{align}
  \psi_0^{\mathrm{cont}}(r) \propto J_0(kr) - \frac{\pi}{2[\gamma +
    \mathrm{ln}(ka/2)]}Y_0(kr),
\label{eq:intwfn}
\end{align}
where $\gamma\approx0.577$ is Euler's constant.

For $E>0$ the scattering phase shift is evaluated using
\eqnref{eq:phaseshift} as
\begin{align}
  \mathrm{cot}\, \delta^\mathrm{cont} (k)  =
  \frac{2}{\pi} \left[\gamma + \mathrm{ln}(ka/2) \right].
\label{eq:analyticscattering}
\end{align}
The pseudopotential must be able to reproduce this phase shift as a function of 
scattering wavevector.

\begin{figure}
\includegraphics[width=\linewidth]{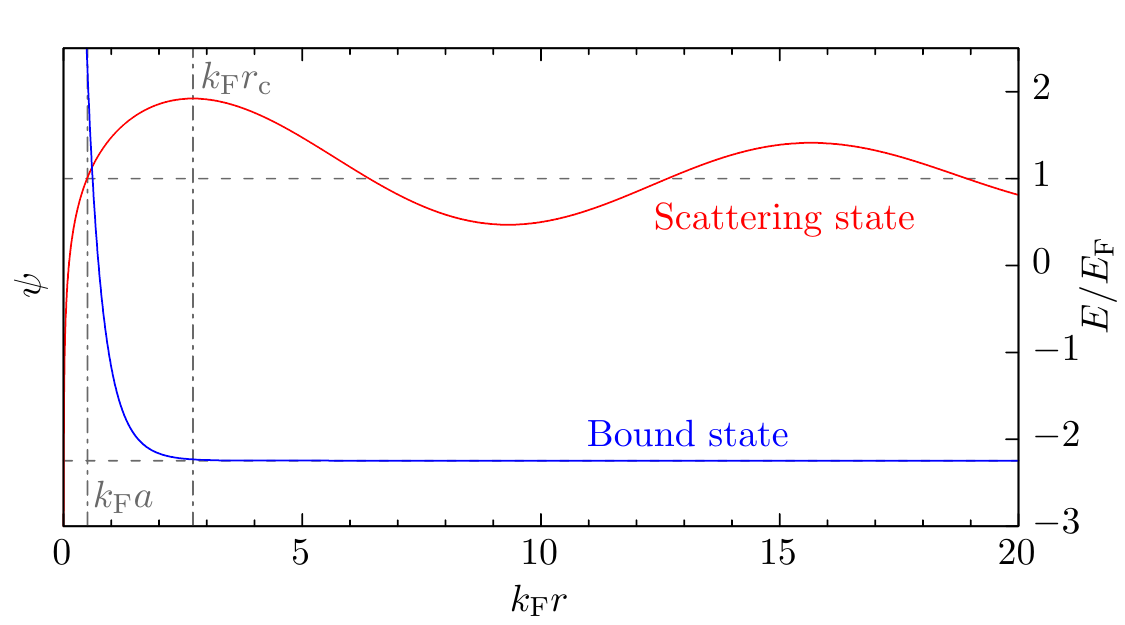}
\caption{(Color online) The bound (blue) and scattering (red) state wavefunctions of
  the contact interaction with $\kf a=1/2$.  The wavefunctions are
  offset by their energies, $E_\mathrm{s}=\Ef$ for the scattering
  state and $E_\mathrm{b}=-\frac{4}{a^2}\exp(-2 \gamma)$ for the bound
  state.  The radius $\rc$ gives the position of the first antinode in
  the scattering state, which is used as the cutoff radius for the
  pseudopotentials.}
\label{fig:wavefunctions}
\end{figure}

For $E<0$ the bound state wavefunction is given by
\eqnref{eq:boundwfn}.  The corresponding bound state energy can be
found from the condition $\scriptB/\scriptA=\complexi$ and
\eqnref{eq:intwfn} as
\begin{align*}
\Eb=-\frac{4}{a^2} \expe^{-2\gamma}.
\end{align*}
Examples of a scattering and bound state wavefunction are shown in
\figref{fig:wavefunctions}.   At large radii the scattering state wavefunction 
takes
the form of a wave, with the first node occurring at $r=a$ in the $k\to 0$ 
limit, whilst the bound state wavefunction decays exponentially. Both 
wavefunctions diverge at particle
coalescence, which makes them difficult to sample in numerical methods.  This 
motivates us to develop a smooth pseudopotential for the contact interaction, 
which will give rise to a wavefunction that is easier to sample numerically.

\section{Derivation of the pseudopotentials}
\label{sec:derivation}

To develop smooth pseudopotentials for the contact interaction we continue to
investigate the two-particle system in a vacuum, where the particles are
distinguishable fermions and an analytical solution exists.  We first focus on 
scattering states where, after reviewing the hard and soft disk potentials that 
are commonly used in ultracold atomic gas calculations, we
construct a pseudopotential using the method proposed by
Troullier and Martins (TM) \cite{Troullier93}.  This method was
originally developed for making pseudopotentials for electron-ion
interactions, but has been successfully applied to other systems of
interacting particles \cite{Bugnion14,Whitehead16}.  Next we construct another, 
``ultratransferable'', pseudopotential (UTP) following the method in 
\reference{Bugnion14}.  We then compare
the accuracy of the TM and UTP pseudopotentials with that of the hard and soft
disk potentials.  Finally, we develop pseudopotentials for bound states.  We 
have made the software used to generate all the pseudopotentials in this 
work available online \cite{MathNb}.

\subsection{Pseudopotentials for scattering states}
\label{sec:ScatteringPseudopotentials}

\subsubsection{Hard disk potential}

Here we briefly review the hard disk potential that is currently used
in many numerical studies of the contact interaction
\cite{Drummond11,Xing90,Bertaina2013}.  The interaction potential has the form
\begin{align*}
V^\mathrm{HD}(r)=\begin{cases}
\infty,& r\le R, \\
0,& r>R,
\end{cases}
\end{align*}
where $R$ is the radius of the potential. Solving the Schr\"odinger
\eqnref{eq:SchrEq} with a boundary condition
$\psi_0^\mathrm{HD}(R)=0$, the wavefunction is given by
\begin{align*}
\psi_0^\mathrm{HD}(r)\propto \begin{cases}
0,&r\le R,\\
-Y_0(kR)J_0(k r)+J_0(kR)Y_0(kr),&r>R,
\end{cases}
\end{align*}
and the scattering phase shift defined by \eqnref{eq:phaseshift} is
\begin{align}
  \mathrm{cot}\, \delta^\mathrm{HD}(k) &=
  \frac{Y_0(kR)}{J_0(kR)}\neweqnline &=\frac{2}{\pi}[ \gamma +
    \mathrm{ln}(kR/2)
  ]\!+\!\frac{(kR)^2}{2\pi}\!+\!\bigO\left((kR)^4\right).
\label{eq:harddiskscattering}
\end{align}
By setting the hard disk radius $R$ equal to the scattering length
$a$, the low energy scattering phase shift from the hard disk has the
same form as the phase shift from the contact potential in
\eqnref{eq:analyticscattering}.  A hard disk potential with
$R=a$ then gives the phase shift for the contact interaction
with an error of order $\bigO\left((ka)^2\right)$, delivering the correct 
scattering properties only in the $k\to 0$ limit.  An example of a
hard disk potential is shown in \figref{fig:potentials}.

\subsubsection{Soft disk potential}

To reduce the error in the scattering phase shift at finite $k$ from that found 
with the hard disk, we may instead use a soft disk potential \cite{Conduit13}
\begin{align*}
V^\mathrm{SD}(r)=\begin{cases}
U,& r\le R, \\
0,& r>R,
\end{cases}
\end{align*}
where $U$ is the height of the soft disk potential. The extra degree
of freedom in this potential relative to the hard disk allows it to remove the 
error in the scattering phase shift in \eqnref{eq:harddiskscattering} of $(k 
R)^2/2\pi$, and so
describe the scattering correct to $\bigO\left((ka)^4\right)$.  We solve the
Schr\"odinger Equation using this potential separately in the regions
$r<R$ and $r>R$, enforcing continuity of the wavefunction and its
derivative at $r=R$, and expanding the scattering phase shift
\eqnref{eq:phaseshift} to second order around $k=0$.  Setting the
first term equal to the contact potential scattering phase shift of
\eqnref{eq:analyticscattering} relates $R$ and the scattering length
$a$ via
\begin{align*}
  R=a_\mathrm{SD}=a \exp \left( \frac{I_0 \left(
        \chi \right)}{\chi I_1 \left(
        \chi \right)} \right),
\end{align*}
where $I_\ell(\chi)$ is the modified Bessel function of the first kind, and the 
factor $\chi^2=U R^2 \approx 2.67$ is obtained
by setting the second order term in the phase shift expansion to zero.
This uniquely specifies a soft disk potential for a given $a$, whose
scattering properties are correct up to order
$\bigO\left((ka)^4\right)$. An example of a soft disk potential is
shown in \figref{fig:potentials}.  It has a larger radius $R$ than the
hard disk potential but a lower height $U$, with the width tending to
zero and the height to infinity as the scattering length goes to zero.

\subsubsection{Troullier--Martins pseudopotential}

\begin{figure}
\includegraphics[width=\linewidth]{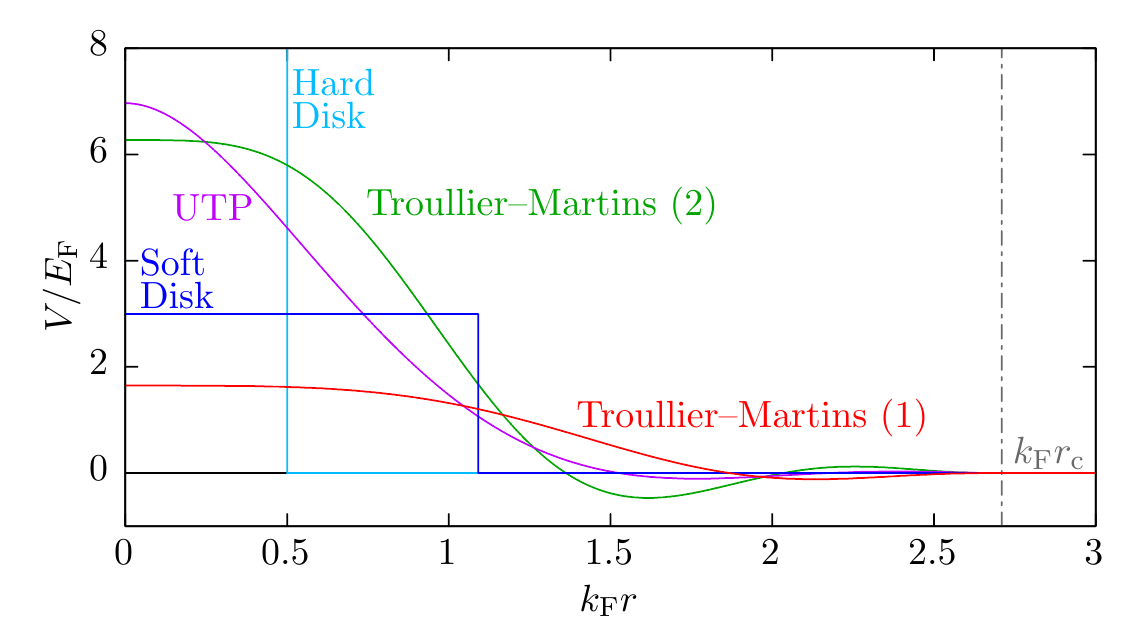}
\caption{(Color online) Scattering state pseudopotentials $V(r)$ for the contact
  potential with $\kf a=1/2$.  The pseudopotential for the hard disc
  with radius $a$ is shown in cyan, the soft disk with radius $a_\mathrm{SD}$ 
in blue, the TM pseudopotentials in red
  and green, and the UTP in magenta.  The pseudopotential cutoff radius $\rc$ 
is the same as in \figref{fig:wavefunctions}.}
\label{fig:potentials}
\end{figure}

The previous subsections showed that the hard and soft disk
potentials give accurate scattering properties only in the
limit of $k\to 0$.  However a Fermi gas contains all the scattering wavevectors 
in the range $0<k \le\kf$, and so the hard and soft disk potentials will give 
rise to inaccurate results.  To demonstrate how the accuracy may be 
improved at finite $k$, we develop
pseudopotentials using the TM formalism \cite{Troullier93,Bugnion14}.
This formalism produces scattering state pseudopotentials that
\begin{enumerate}
\item should reproduce the phase shift of the contact potential accurately
  for all scattering wavevectors in the Fermi sea;
\item are smooth everywhere, which accelerates numerical calculations;
  and
\item for repulsive interactions do not support a bound state.
\end{enumerate}

This formalism requires two prescribed parameters, namely the
calibration wavevector $\kc$ at which the resulting
pseudo-wavefunction has identical scattering properties to the contact
potential, and a cutoff radius $\rc$ at which the pseudopotential
smoothly becomes zero.

The calibration wavevector must be chosen for each system.  For
example, in a superfluid we might choose $\kc=\kf$, as that is where
the most important physics of Cooper pair formation occurs.  For a fermionic 
gas we choose
$\kc=\kf/2$, which minimizes the average phase shift error
\cite{Whitehead16}.

By choosing the cutoff radius to be larger than the radius of the first node in 
the analytic wavefunction, which is at $r\approx a$ in 
\figref{fig:wavefunctions}, we 
ensure that the pseudo-wavefunction does not contain the innermost node that 
corresponds to the bound state of the contact interaction \cite{Bugnion14}.  In 
order to avoid unnecessarily removing scattering states from the potential the 
cutoff radius must also be smaller than the radius of the second node, and so 
we choose the cutoff radius to be at the first antinode of the
wavefunction with $k=\kc$, shown in \figref{fig:wavefunctions}. 

The TM pseudopotential takes the form
\begin{align*}
V^\mathrm{TM}(r)=\begin{cases}
\kc^2+p''+p'^2+\frac{p'}{r},&r\le\rc,\\
0,&r>\rc,
\end{cases}
\end{align*}
where the polynomial $p(r)=\sum_{i=0}^6 c_i r^{2i}$, and primes
indicate derivatives with respect to $r$.  The coefficients $\{c_i\}$
are determined by a set of constraints on the pseudopotential and
pseudo-wavefunction, whose form is \mbox{$\psi(r)=\exp [p(r)]\,$}: that the
pseudo-wavefunction is smooth up to the fourth derivative at $\rc$;
that the pseudopotential has zero curvature at the origin; and that
the norm of the pseudo-wavefunction within $\rc$ equals that of the
wavefunction from the real contact potential 
\cite{Troullier93,Bugnion14,Whitehead16}.  This gives rise to a
set of coupled equations for the $\{c_i\}$ of which one is quadratic
and the others linear: there are therefore two separate branches of
solutions, which give rise to two separate TM pseudopotentials.

In \figref{fig:potentials} we compare all of the discussed pseudopotentials for 
the contact interaction, with $\kf a=1/2$.  The TM
pseudopotentials, being everywhere smooth and finite, are easier to work
with numerically than the hard and soft disks, and they do not
introduce discontinuities in the first derivative of the wavefunction.  The 
potential labeled Troullier--Martins (1) in \figref{fig:potentials} is smaller 
than Troullier--Martins (2) at particle coalescence, but larger at further 
separations to give similar average scattering. 

\subsubsection{Errors in scattering phase shift}

\begin{figure}
\includegraphics[width=\linewidth]{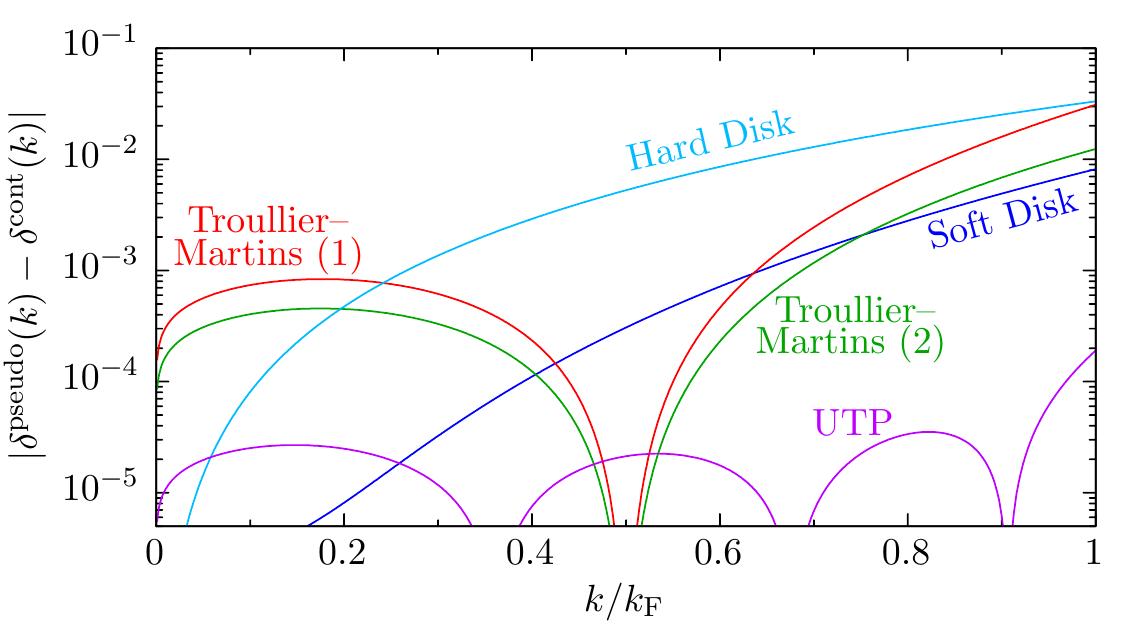}
\caption{(Color online) The error in the scattering phase shift
  \mbox{$|\delta^\mathrm{pseudo}(k)-\delta^\mathrm{cont}(k)|$}
  as a function of scattering wavevector for the different
  pseudopotentials at $\kf a=1/2$. The error from the hard disk is shown in 
cyan, the error from the soft disk in blue, the errors from the two TM 
pseudopotentials in red and green, and the error from the UTP in magenta.}
\label{fig:phase_shift}
\end{figure}

The quality of a pseudopotential for scattering states
may be determined by how accurately it reproduces the phase shift of
the contact potential.  All information
on the difference between the pseudopotential and contact potential
can be obtained from the wavefunction just beyond the edge of the
pseudopotential.  We match the analytical pseudo-wavefunction that solves 
\eqnref{eq:SchrEq}, $\psi$, and its first derivative to the non-interacting 
solution \eqnref{eq:nonintsolution} at a radius $\re$ beyond the radius of the 
pseudopotential.  This leads to an expression for the scattering phase shift
\begin{align*}
\mathrm{cot}\, \delta(k) = \frac{\frac{\psi'(\re)}{\psi(\re)}Y_0(k \re)+k Y_1(k 
\re)}{\frac{\psi'(\re)}{\psi(\re)}J_0(k \re)+k J_1(k \re)}.
\end{align*}

We calculate the difference in the phase shift between the contact interaction 
and
pseudopotentials, showing the error in the calculated phase shifts from using 
the
pseudopotentials \mbox{$|\delta^\mathrm{pseudo}(k)-\delta^\mathrm{cont}(k)|$} 
in \figref{fig:phase_shift}, with $\re=\rc$.  The hard and soft disk potentials 
are exact in the limit of $k\to 0$, but deviate away from that point, with the 
soft disk performing better than the hard disk.  The TM (1)
pseudopotential is on average around twice as accurate as
the hard disk potential, with the TM (2) pseudopotential being
around twice as accurate again, and the soft disk being
another $1.3$ times more accurate.  Both TM pseudopotentials capture
the scattering behavior perfectly at $\kc=\kf/2$ but deviate at all other 
scattering wavevectors, which is a consequence of the norm-conserving condition 
on the
pseudo-wavefunctions.  To further improve the accuracy of the pseudopotentials, 
a natural extension to the formalism is to construct a pseudopotential that 
minimizes this deviation in the phase shift over all wavevectors $k\le\kf$.  We 
propose such a pseudopotential here, referring to it as an ``ultratransferable 
pseudopotential'' (UTP) \cite{Bugnion14}.

\subsubsection{Ultratransferable pseudopotential}

Similarly to the TM pseudopotential, the UTP takes a polynomial form within a 
cutoff radius $\rc$,
\begin{align*}
V^\mathrm{UTP\!}(r)\!=\!\begin{cases}
\!\left(\!1\!-\!\frac{r}{\rc}\!\right)^2 \!\left[ u_1\!\left( 1+\frac{2r}{\rc} 
\right)\! +\! \displaystyle\sum_{i=2}^{N_u} u_i 
\left(\!\frac{r}{\rc}\!\right)^{\!i} \right]\!,&\!\!\!\!r\le\rc,\\
0,&\!\!\!\!r>\rc,
\end{cases}
\end{align*}
with $N_u=3$.  The term $(1-r/\rc)^2$ ensures that the pseudopotential goes 
smoothly to zero at $r=\rc$, and the component $u_1(1+2r/\rc)$ constrains the 
pseudopotential to have zero derivative at the origin.  This ensures that the 
pseudo-wavefunction is smooth, easing the application of numerical methods.

To determine the coefficients $\{u_i\}$ we numerically solve the scattering 
problem, extract the scattering phase shift $\delta^\mathrm{UTP}(k)$, and then 
minimize the total squared error in the phase shift over all scattering 
wavevectors $k$,
\begin{align*}
\langle \left| \delta^\mathrm{UTP}(k) - \delta^\mathrm{cont}(k) \right|^2 
\rangle&= \\
\int_0^{\kf}&\left| \delta^\mathrm{UTP}(k) - \delta^\mathrm{cont}(k) \right|^2 
g(k/\kf) \dd k,
\end{align*}
where the weighting is given by the density of states in the center of mass
frame \mbox{$g(k)=k( 
4-\frac{8}{\pi}[ k \sqrt{1-k^2} + \arcsin(k) ] )$} \cite{Whitehead16}. An 
example UTP is shown in \figref{fig:potentials}, confirming that this 
construction gives smooth potentials.  The scattering phase shift error from 
the UTP is shown in \figref{fig:phase_shift}, demonstrating that the UTP 
construction creates pseudopotentials that are significantly more accurate than 
the Troullier--Martins pseudopotentials and soft disk potential, and some $200$ 
times more accurate than the hard disk.  This is achieved by the phase shift 
error from the UTP being optimized to be zero at three different wavevectors, 
as opposed to the single wavevector for the TM pseudopotentials.

\subsection{Pseudopotentials for bound states}

Pseudopotentials may also be constructed for particles in a bound
state, with $E<0$.  In order to accurately imitate the
contact potential, the pseudopotentials must reproduce the bound state
energy of the contact potential $\Eb=-(4/a^2)\exp(-2\gamma)$, and also
must accommodate only one bound state.  We first discuss the square
well pseudopotential, which has been used in previous ultracold atomic
gas calculations, and then again develop smooth pseudopotentials using
the TM formalism. For bound states there is no quantity like the
scattering phase shift that can be used to directly determine the
quality of the pseudopotentials.  We therefore demonstrate their
accuracy in a two-body inhomogeneous system in \secref{sec:harmonic}.

\subsubsection{Square well potential}

The square well potential has the form
\begin{align*}
V^\mathrm{SW}(r)=\begin{cases}
-U,& r\le R, \\
0,& r>R.
\end{cases}
\end{align*}
This potential may be made arbitrarily close to the bound state
contact interaction by taking the well radius $R\to 0$ and depth $U\to
\infty$.  Decreasing $R$, however, reduces the sampling efficiency and
thereby increases the computational cost.  We require $R$ to be less 
than the average interparticle separation $\sim 1/\kf$, in order to 
avoid the unphysical situation of three or more particles interacting 
simultaneously.
In \secref{sec:harmonic} we investigate the
$R$ dependence of the accuracy of the square well pseudopotential.

Because there is no analogue of the scattering phase shift for the
bound system it is not possible to uniquely define a $U$ and $R$ for a
given $a$, as we did for the soft disk potential in which $U$
and $R$ were related by the second order term in the expansion.
However one parameter may be determined by ensuring that the bound
state energy of the potential is $\Eb$, and for a given $R$ the value
of $U$ this sets can be found as a solution to
\begin{align*}
  \frac{-J_0(k_1 R)}{k_1 J_1 (k_1 R)} = \frac{J_0(i k_2 R) - i Y_0(-i
    k_2 R)}{k_2 \left( -i J_1(i k_2 R) + Y_1(-i k_2 R) \right)},
\end{align*}
where $k_1=\sqrt{U-|\Eb|}$ and $k_2=\sqrt{|\Eb|}$.  An example of a square well 
potential is shown in \figref{fig:bound_potentials}.  Except in the limit of 
being infinitely deep and narrow, the square well potential does not give rise 
to the same wavefunction as the true contact interaction, but within the 
potential the wavefunction and therefore probability density is too small.  
This means that in the presence of an external potential (for example an 
harmonic trap, as in \secref{sec:harmonic}) there is too much weight at large 
particle separations, giving rise to inaccurate values of the system's energy.  
As $R \to 0$ the wavefunction approaches the exact form given by 
\eqnref{eq:boundwfn}.

\subsubsection{Troullier--Martins pseudopotential}
\label{sec:TMpseudo}

\begin{figure}
\includegraphics[width=\linewidth]{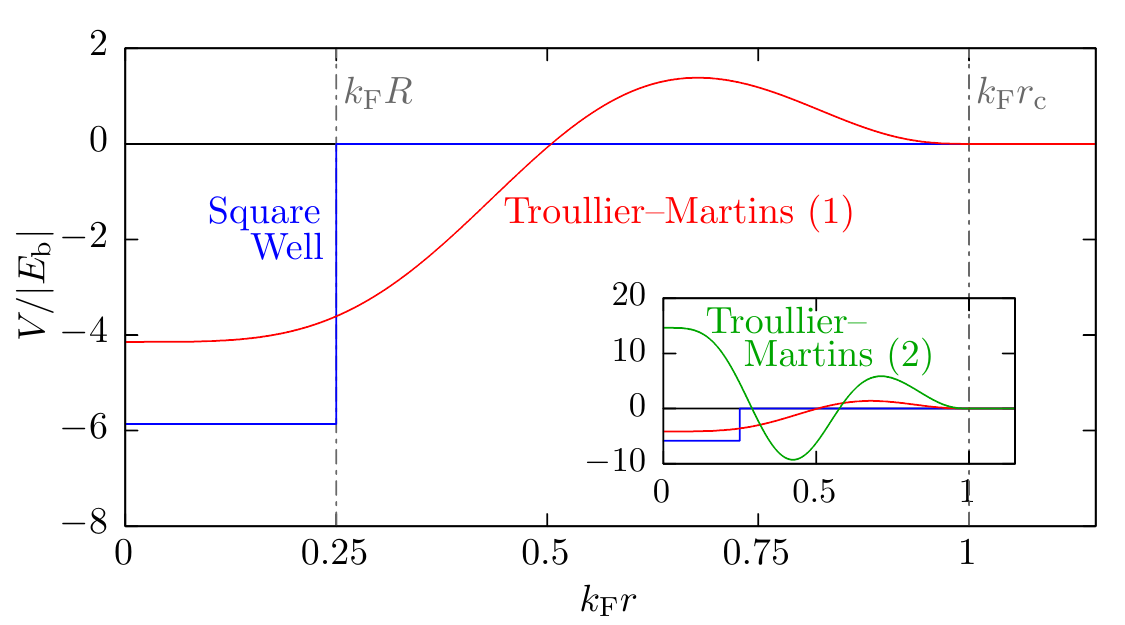}
\caption{(Color online) Bound state pseudopotentials for the contact potential with
  $\kf a=1/2$.  A square well with $\kf R=1/4$ is shown in blue, and
  the TM (1) pseudopotentials with $\kf \rc=1$ is shown in red.
  Inset: the TM (2) pseudopotential, shown in green and with $\kf \rc=1$, 
behaves qualitatively
  differently near particle coalescence.}
\label{fig:bound_potentials}
\end{figure}

The Troullier--Martins pseudopotential resolves the problem of having too much 
weight at large particle separations by being a norm-conserving 
pseudopotential, and so has the correct amount of weight within and outside of 
its cutoff radius. 
The construction of the TM pseudopotential for the bound state is
identical to that of the scattering state, except that the calibration
energy is now given by the bound state energy $\Ec=\Eb$.  The cutoff
radius $\rc$ should be kept smaller than the average interparticle
separation $\sim 1/\kf$ to reduce the probability of three or more
particles interacting at once, but there is no lower bound on $\rc$:
similarly to the case of the square well, reducing $\rc$ increases the
accuracy but also the computational cost of simulations.  The
square well and TM pseudopotentials are shown in
\figref{fig:bound_potentials}.

One of the TM pseudopotentials, labeled (1) in
\figref{fig:bound_potentials}, behaves as would be expected
qualitatively for a short-ranged potential giving rise to a bound
state: it has a large negative region near particle coalescence.  The
other TM solution, labeled (2) and shown in the inset to
\figref{fig:bound_potentials}, does not show this behavior, instead
having an attractive region at finite particle separation.  This will
give rise to a non-zero expected separation between bound particles,
which is physically discordant with the contact interaction.  We therefore 
reject
the TM (2) pseudopotential because of its unphysical behavior and select the
TM (1) pseudopotential instead, referring to
it henceforth simply as the TM pseudopotential.

Since all particles in bound states have approximately the same energy, the UTP 
formalism does not offer any advantage in this system.  We now move on to 
testing the pseudopotentials in an inhomogeneous two-body system.

\section{Two fermions in an harmonic trap}
\label{sec:harmonic}

We have constructed pseudopotentials that describe the scattering
behavior of two isolated fermions.  To test the pseudopotentials we
turn to the experimentally realizable \cite{Murmann15,Zurn12} system
of two distinguishable fermions in a circular harmonic trapping
potential $\frac{1}{4}\omega^2 r^2$ of frequency $\omega$.  This
system also has the advantage of being analytically soluble, which
provides a stringent test for the pseudopotentials that we will use in
many-body simulations.

\subsection{Analytic energy levels}

\begin{figure}
\includegraphics[width=\linewidth]{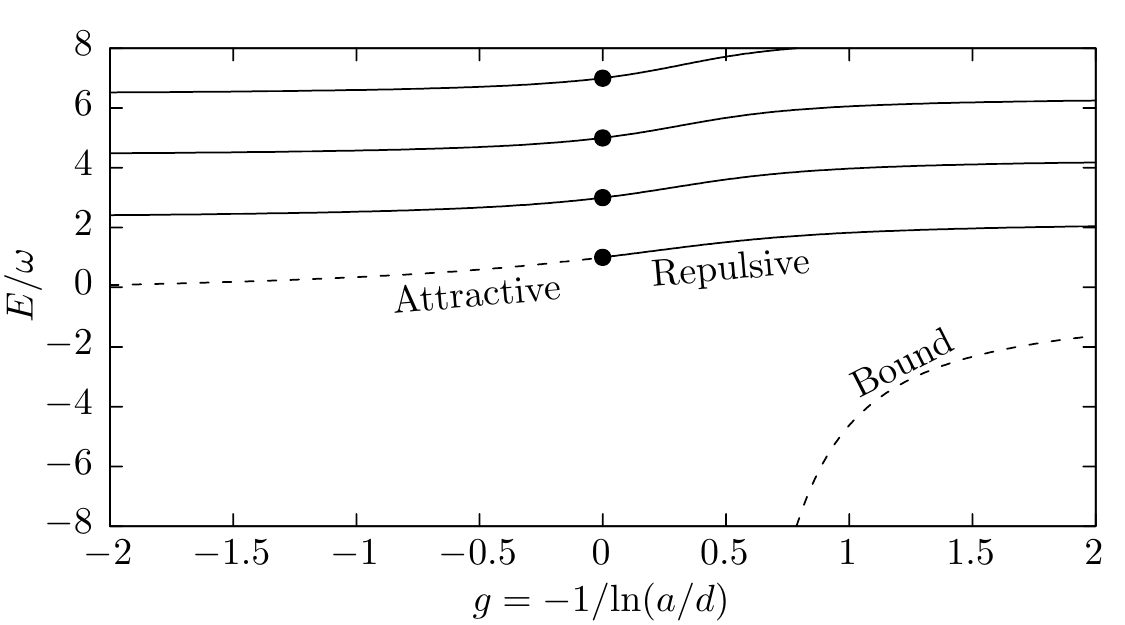}
\caption{Analytic energy levels for two particles in an harmonic trap
  as a function of the dimensionless interaction strength
  $g=-1/\ln\left(a/d\right)$.  The excited states (solid lines)
  correspond to the scattering states of the contact potential and the
  ground state (dashed line) corresponds to the bound state.  The
  non-interacting energies are shown by circles along the line $g=0$.}
\label{fig:harmonic_energies}
\end{figure}

In the center-of-mass frame the Schr\"odinger Equation for two
distinguishable fermions in an harmonic trap is given by
\begin{align*}
-\nabla^2 \psi(r)+\frac{1}{4}\omega^2 r^2 \psi(r)+\Vcont(r)\psi(r)= E\psi(r)
\end{align*}
where the interparticle interaction term $\Vcont(r) \psi$ can be replaced by a 
boundary condition given
by \eqnref{eq:boundary}.  For the contact interaction the energy levels in the 
center-of-mass frame are solutions to the nonlinear equation 
\cite{Liu10,Busch98}
\begin{align}
  \Psi\left(-\frac{E}{2\omega}+\frac{1}{2}\right) =
  \ln\left(\frac{d^2}{a^2}\expe^{-2\gamma}\right),
\label{eq:harm_energies}
\end{align}
where $d=\sqrt{2/\omega}$ is the characteristic length scale of the
trap and $\Psi$ is the digamma function.  These solutions are shown in
\figref{fig:harmonic_energies} as a function of the dimensionless 
interaction strength $g=-1/\ln(a/d)$. In the non-interacting case $g=a=0$
the energies have the expected values $E=\omega(2n+1)$ for
non-interacting particles.  As the repulsive interaction strength $g>0$ in
\figref{fig:harmonic_energies} increases, the energy increases and at
 $g\to\infty$ joins onto the energy of the attractive branch at 
$g\to -\infty$, in an analogue of unitarity in the BEC-BCS 
crossover \cite{Ries15}.  The bound state of the contact potential 
survives in this inhomogeneous system as the deep bound state at $g>0$.

\subsection{Accuracy of the pseudopotentials}

\begin{figure}
\includegraphics[width=\linewidth]{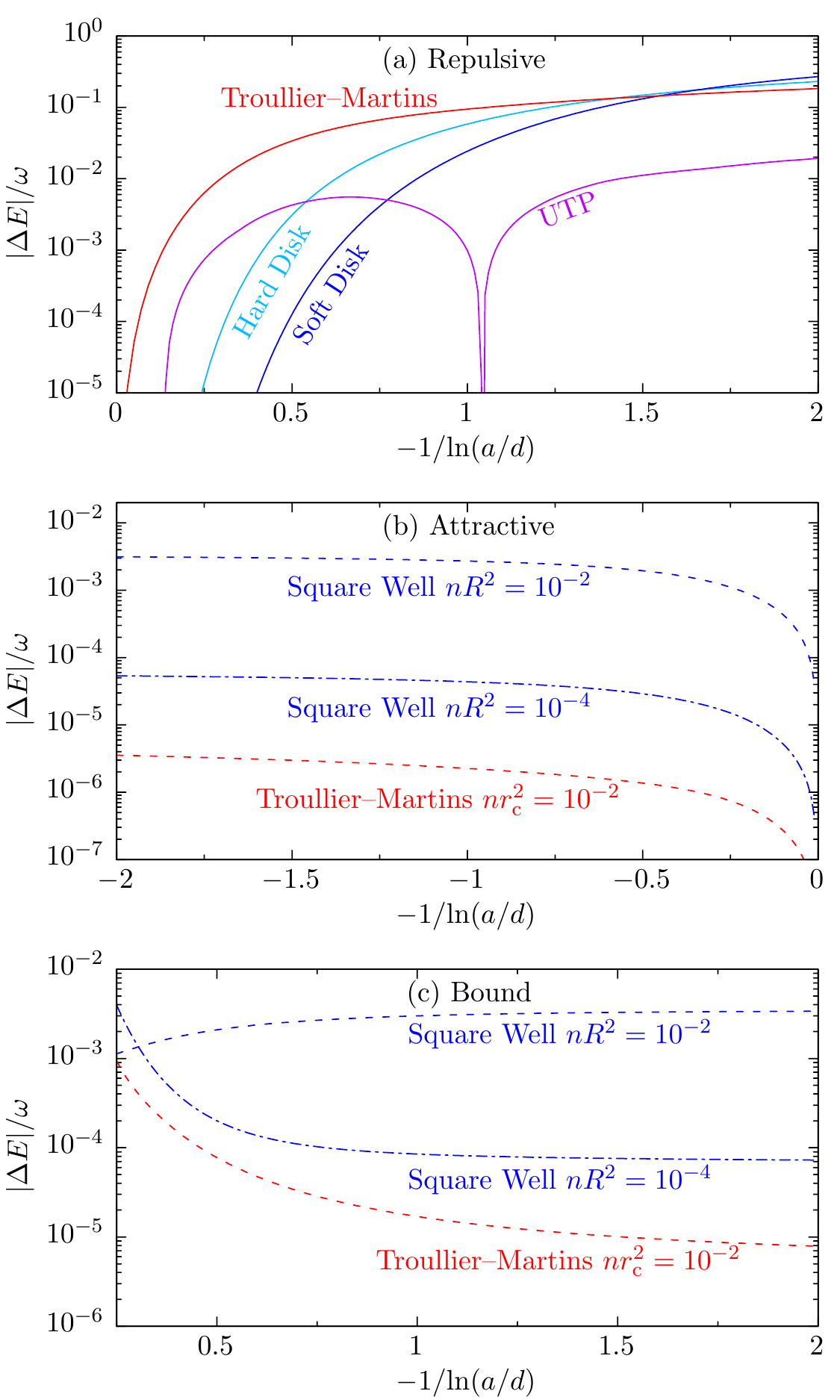}
\caption{(Color online) Error in the center-of-mass energy of two fermions in an harmonic trap
  calculated using pseudopotentials from the analytic value of the
  energy from \eqnref{eq:harm_energies}.
 (a) Error in the center-of-mass energy of particles with a repulsive
  interaction as a function of interaction strength for the hard disk in cyan, 
the soft disk in blue, the TM pseudopotential in red, and the UTP in magenta. 
(b) Error in the center-of-mass energy
  of particles with a weakly attractive interaction as a function of
  interaction strength. The square well pseudopotentials have radii given by $n
  R^2=10^{-4}$ and $n R^2=10^{-2}$, and the TM pseudopotential has a
  cutoff radius given by $n \rc^2=10^{-2}$, with the different cutoff radii 
denoted by different types of line dashing. (c) Error in the bound state
  energy as a function of interaction strength.}
\label{fig:harmonic_errors}
\end{figure}

We compare the estimates of the center-of-mass energies of two particles in an
harmonic trap to the analytic result in:
\figref{fig:harmonic_errors}(a), for repulsive interactions;
\figref{fig:harmonic_errors}(b) for attractive interactions; and
\figref{fig:harmonic_errors}(c) for bound particles.

In the repulsive case, we find that the hard and soft disk potentials and TM 
pseudopotential
are accurate at small interaction strengths, but at large interaction
strengths the error in the calculated energies is greater than 10\%.
The UTP pseudopotential is around 10 times more accurate at high
interaction strengths, and becomes exact in the non-interacting
limit.

To choose the radii of the potentials for the attractive and bound
branches we follow the approach used in
\reference{Bugnion14} and use a TM pseudopotential with a
cutoff radius of $n \rc^2=10^{-2}$, where $n=\omega/2\pi$ is the peak
density of two non-interacting particles in the trap.  We compare this
to square wells with radii given by the same $n R^2=10^{-2}$ and the smaller $n 
R^2=10^{-4}$ \cite{Bugnion14,Astrakharchik04}.  We note that in both the 
attractive and bound branches, reducing the well radius increases the accuracy 
of the square well potential, but that the TM pseudopotential gives up to 10
times higher 
accuracy than a square well with a radius $1/10$ the size.  The ability to use 
a larger cutoff radius with the TM pseudopotential brings significant
benefits in numerical sampling of the potential, with the sampling efficiency 
expected to scale as $\sim \rc^2$.  The increased accuracy can 
be related to the fact that the square well gives rise to wavefunctions with 
too much weight at large particle separations, raising the energy in the 
external trap, whilst the TM pseudopotential is norm-conserving, having the 
correct weight in the wavefunction outside $\rc$.  
The norm-conservation condition
ensures that the
TM pseudopotential gives a bound state wavefunction that is robust against 
changes in the
local environment, and hence performs well in the spatially varying harmonic 
trap.  As opposed to 
the single calibration energy of the TM pseudopotential, in constructing the 
UTP we would average over 
a range of energies.  This would offer no advantage in the 
attractive and bound branches, where there is a definite binding energy for 
the pair 
of particles, and so we do not examine the UTP in these branches.

We have shown that for particles in an harmonic trap with attractive 
interactions, the TM pseudopotential gives an increase in both accuracy and 
sampling efficiency relative to the square well potential. For two particles 
with repulsive interactions, the use of a UTP can offer a 10-fold increase in 
accuracy relative to using the TM pseudopotential or hard or soft disk 
potentials.  We now go on to demonstrate the scaling benefits of the UTP in a 
many-body simulation.

\section{Fermi gas}
\label{sec:fermi}

Having demonstrated the effectiveness of the UTP for studying the two
body scattering problem and two distinguishable fermions in an harmonic
trap, we now demonstrate the advantages of the UTP in a prototypical
setting: a two-dimensional homogeneous Fermi gas. Such a system serves
as a benchmark for cold atom experiments
\cite{Frohlich2011,Martiyanov2010} and also as a model for electrons
in conductors.

We focus on the repulsive branch of the contact interaction. Here the hard and 
soft disk potentials are uniquely defined for a given interaction strength, and 
may not be improved to attain arbitrarily high accuracy, as is possible in the 
attractive and bound branches by reducing the well radius to zero.  This allows 
us to demonstrate the intrinsic benefits of the UTP formalism over the hard and 
soft disk potentials.

The smoothness of the UTP relative to the hard and soft disk potentials will be
reflected in the many-body wavefunction, which will make it easy to
work with numerically.  Having shown in \secref{sec:harmonic} that the
UTP is more accurate than the competing hard and
soft disk potentials and Troullier--Martins pseudopotential, we proceed here to 
verify 
the accuracy of the UTP by comparing the energy of a Fermi gas with
first- and second-order perturbation theory calculations
\cite{Engelbrecht1992,Engelbrecht1992b,He2014}.

\subsection{Formalism}

To calculate the ground state energies we use the diffusion Monte Carlo
(DMC) technique. DMC is a highly-accurate Green's function projector
method for determining ground state energies and expectation values
\cite{Umrigar1993,Ceperley1980,Foulkes2001}, and it is well-suited to
investigating homogeneous gaseous phases. We use the \textsc{casino}
implementation \cite{Needs2010} of the DMC method with a
Slater--Jastrow trial wavefunction $\Psi = \expe^J D_{\uparrow}
D_{\downarrow}$, where $D_{\uparrow}$ ($D_{\downarrow}$) is a Slater
determinant of plane-wave states for the spin up (down) channel. The
Jastrow factor $\expe^J$ describes correlations between particles,
with
\begin{align}
J= \!\!\!\sum_{\substack{
      j \neq i \\
      \alpha,\beta \in \{\uparrow,\downarrow\}
    }} \!\!\left( 1- \frac{r_{ij}}{L_\mathrm{c}} \right)^3 u_{\alpha 
\beta}\left(r_{ij}\right) \Theta\left(L_\mathrm{c} - r_{ij}\right),
\label{eq:jastrow}
\end{align}
where $r_{ij}=|\vect{r}_i-\vect{r}_j|$ is the distance between two
particles with labels $i$ and $j$, and $u_{\alpha\beta}$ are
eighth-order polynomials, whose parameters are optimized using
variational Monte Carlo subject to the symmetry requirements
$u_{\uparrow\uparrow} = u_{\downarrow\downarrow}$ and
$u_{\uparrow\downarrow} = u_{\downarrow\uparrow}$. $L_\mathrm{c}$ is
a cutoff length that we set equal to the radius of a circle inscribed within 
the simulation cell, and $\Theta$ is the Heaviside step function.

We calculate the ground state energy expectation value for 49 spin-up
and 49 spin-down particles in a homogenous two-dimensional system for
increasing interaction strengths $-1/\ln(\kf a)$ up to a maximum value
of 1.8 before the system would phase separate into a fully polarized state. To
accurately capture the hard disk wavefunction at small inter-particle
distances in our DMC simulations we add an additional term to
the Jastrow factor in \eqnref{eq:jastrow},
\begin{align}
u_\mathrm{H}(r)=\begin{cases}
-\infty, & r \leq R, \\
\log[\tanh(\frac{r/R-1}{1-r/L_\mathrm{c}})], & R < r < L_\mathrm{c}, \\
0, & r \geq L_\mathrm{c}, \\
\end{cases}
\label{eq:jastrow2}
\end{align}
as in \reference{Drummond11}, where $R$ is the hard disk radius. In
the present study the additional term applies to opposite spins
only. 

We extrapolate to zero DMC timestep to obtain accurate ground
state energies. For each data point we run three simulations with
timesteps $0.25 dt, 0.5 dt, dt$, \cite{Lee2011} with $dt$ the maximum
timestep in the linear regime, and extrapolate to zero timestep by
minimizing the weighted least squares fit.  All error bars represent the DMC 
stochastic error combined with the concomitant uncertainty in the timestep 
extrapolation. We expect that the use of a quadratic DMC algorithm would give 
similar results \cite{Mella00,Sarsa02}. 

\begin{figure}
  \includegraphics[width=\linewidth]{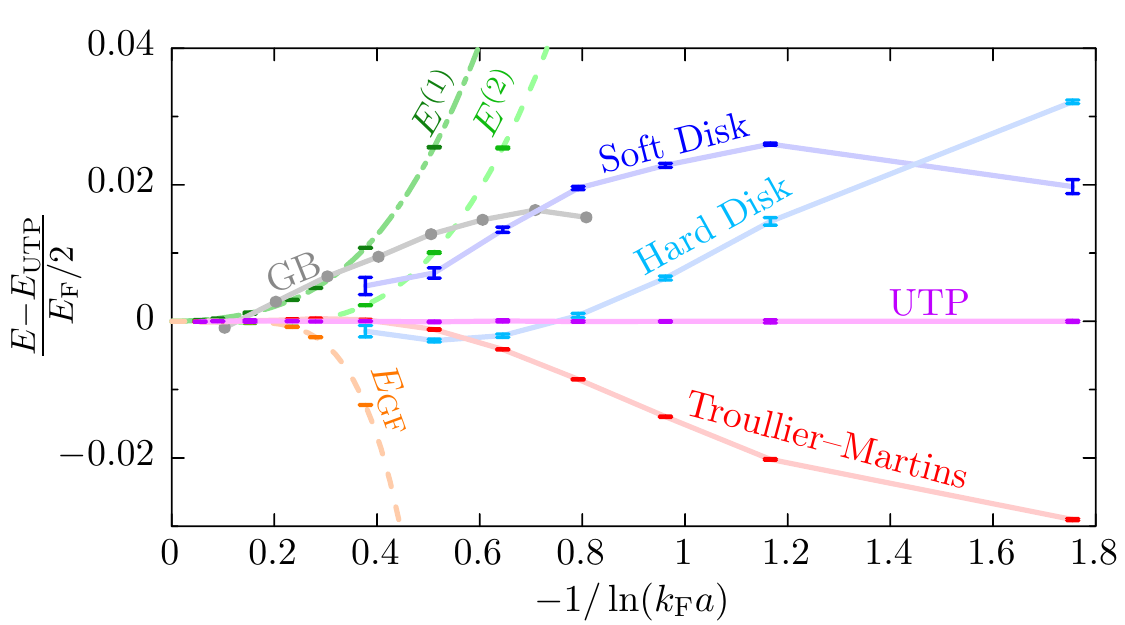}
  \includegraphics[width=\linewidth]{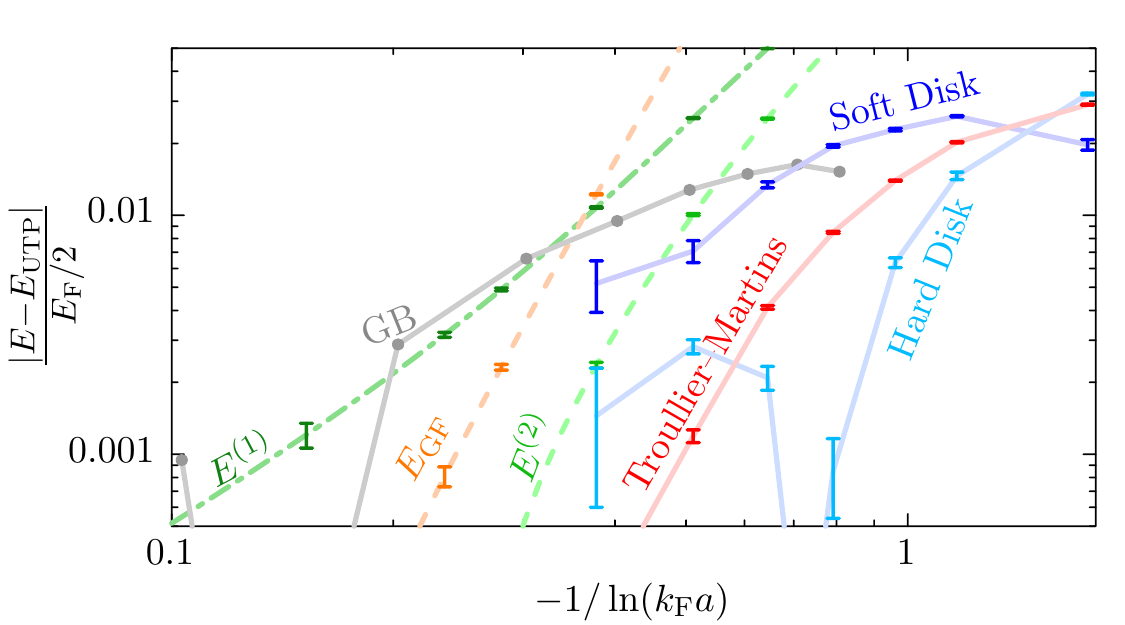}
  \caption{(Color online) (Top) Differences in ground state energy from the result
    obtained with the UTP as a function of interaction
    strength, normalized by the energy of the non-interacting
    system. The green lines denoted $E^{(1)}$ and $E^{(2)}$ are predictions
    from first- and second-order perturbation theory
    \cite{Engelbrecht1992,Engelbrecht1992b,He2014} and $E_\mathrm{GF}$ is the
    result of a Galitskii-Feynman partial resummation of Feynman
    diagrams reported in \reference{He2014}, shown in orange. GB is the Monte 
Carlo result
    from \reference{Bertaina2013}, calculated using a hard disk
    potential and shown in gray, and our results using the hard disk are shown 
in cyan, the soft disk in blue, the TM pseudopotential in red, and the UTP in 
magenta.   (Bottom) The same results on a logarithmic scale.}
  \label{fig:fermi_gas_errors}
\end{figure}

\subsection{Results}

In \figref{fig:fermi_gas_errors} we compare ground state energies of
the Fermi gas obtained using the different potentials. It is clear that
for $-1/\ln (\kf a) > 0.7$ both the hard and soft disk potentials, as
well as the Troullier--Martins pseudopotential, are insufficient to
obtain the desired $10^{-4}\Ef$ accuracy that has been obtained in
other DMC studies of homogeneous systems
\cite{Bugnion14,Whitehead16,Lloyd-Williams2015,Conduit09}.

To verify the DMC results we compare our estimates for the ground
state energy with perturbation theory
\cite{Engelbrecht1992,Engelbrecht1992b,He2014}. As can be seen in
\figref{fig:fermi_gas_errors}(b), first order perturbation theory
$E^{(1)}=\frac{\Ef}{2}(1+[-1/\ln(\kf a)])$ deviates quadratically in
the interaction strength $-1/\ln(\kf a)$ from the UTP result as
expected, and second order perturbation theory
\begin{equation*}
  E^{(2)} = \frac{\Ef}{2} \bigg[1+ \Big(\frac{-1}{\ln (\kf a)}\Big) + 
\Big(\frac{3}{4} - \ln (4 \mathrm{e}^{-\gamma})\Big) \Big(\frac{-1}{\ln (\kf 
a)}\Big)^2 \bigg]
\end{equation*}
deviates cubically in $-1/\ln(\kf a)$ and
outperforms first order perturbation theory. In
\figref{fig:fermi_gas_errors} we also show the result obtained in
\reference{He2014} using a partial resummation of Feynman diagrams in
the Galitskii-Feynman (GF) scheme which is correct to order
 $\mathcal{O}\left( [-1/\ln(\kf a)]^3 \right)$, and note that this indeed 
deviates cubically in interaction strength from the UTP result.  The agreement 
of the scaling behavior of the energy calculated using the UTP with interaction 
strength when compared to these analytic results confirms the accuracy of the 
UTP.

In addition to the analytic approximations, we compare our DMC results
with an independent study using the hard disk potential and the same
number of particles in \reference{Bertaina2013}, labeled GB. We note that their 
predicted energies are higher than those from our DMC calculations using the 
hard disk potential, and as DMC is a variational method this indicates that our 
trial wavefunction is likely more accurate than was available to the authors of 
\reference{Bertaina2013}, possibly due to our inclusion of a Jastrow factor 
with variational parameters.

\begin{figure}
  \includegraphics[width=\linewidth]{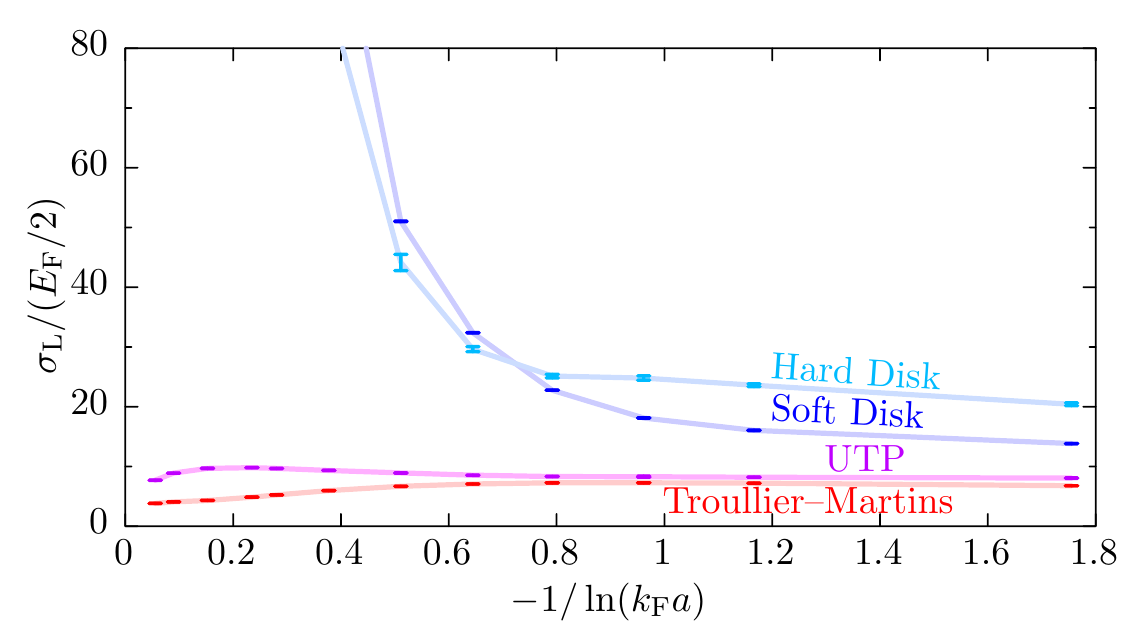}
  \caption{(Color online) Standard deviation of the local energy distribution
    of the trial wavefunction. The hard
    and soft disk pseudopotentials exhibit a larger standard
    deviation, due to the sudden changes in energy when two particles
    approach one another.}
  \label{fig:fermi_gas_var}
\end{figure}

Having confirmed the accuracy of the UTP we now examine its performance 
benefits.  The local energy, $E_\mathrm{L} = \Psi^{-1}\hat{H}\Psi$, is a crucial
quantity in DMC calculations \cite{Needs2010}.  The stochastic error
in a DMC calculation is proportional to the standard deviation
$\sigma_\mathrm{L}$ in the local energy distribution, and therefore a smoother
local energy will give rise to more accurate results for the same
computation time.  \figref{fig:fermi_gas_var} shows the standard
deviation of the local energy distribution of the trial wave
function when using all of our pseudopotentials. Both the UTP and TM 
pseudopotentials benefit from their
smoothness in obtaining a lower local energy standard deviation
compared to the hard and soft disk potentials.  For weak interactions
the hard disk potential benefits from an additional Jastrow factor
term, \eqnref{eq:jastrow2}, relative to the soft disk potential, whose height 
$U$ also diverges as $a\to 0$.  However for
larger interactions the soft disk potential results in a smoother
wavefunction than the hard disk potential and therefore has lower
local energy variance. The variance in the local energies diverges for
the hard and soft disk potentials for weak interactions, whereas it
decays for the UTP and TM pseudopotentials.  The standard deviation for the TM
pseudopotential is slightly lower than the UTP at all interaction strengths, 
which is understood
from the larger size of the potential for the UTP in \figref{fig:potentials} 
compared
to the TM pseudopotential.  This behavior is similar
to the 3D case reported in \reference{Bugnion14}.

The reduced variance in the local energy lowers the computational effort $T$ 
required for a DMC calculation, which scales as
$T \propto \sigma_{\mathrm{L}}^2/dt$
\cite{Lloyd-Williams2015,Whitehead16,Ma05}.  From
\figref{fig:fermi_gas_var} we see that at intermediate interaction
strength $-1/\ln (\kf a) = 0.8$ the variance of the local energy for
the UTP is 2.7 and 3.0 times lower than for the soft and hard disk
potentials respectively, corresponding to a speedup of 7.5 and 9.1.

\begin{figure}
  \includegraphics[width=\linewidth]{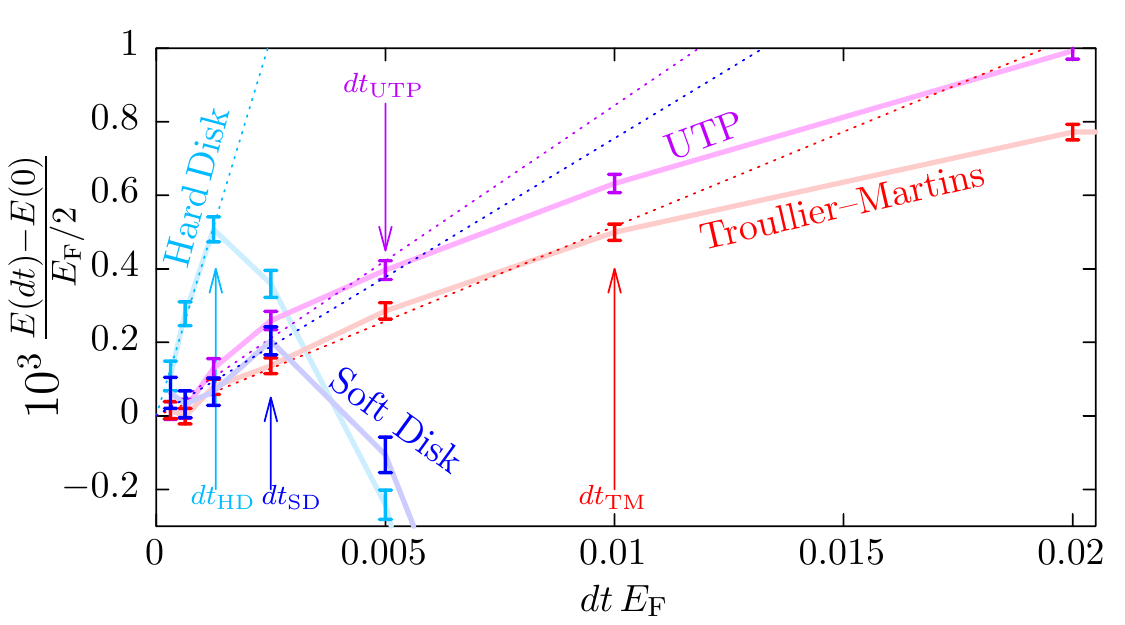}
  \caption{(Color online) Error in the estimated ground state energy as a function of
    DMC timestep. The results using the hard disk are shown in cyan, the soft 
disk in blue, the TM potential in red and the UTP in magenta, with solid lines 
indicating the values calculated using DMC and dotted lines a linear 
extrapolation.  This enables the identification of when the error leaves the 
linear regime.}
  \label{fig:timesteps}
\end{figure}

In addition to the lower local energy variance, our 
pseudopotentials offer an additional speedup.  The DMC estimate of the energy 
must be extrapolated to zero timestep, and the larger the region of linear 
dependence of energy on timestep, the larger timestep can be used.  This 
reduces computational effort even further, as $T\propto 1/dt$.  In 
\figref{fig:timesteps} we observe that the extent of the linear regime of the 
error in ground state energy with timestep differs between the 
pseudopotentials: it extends up to \mbox{$dt_\mathrm{HD} = 1.25 \times 
10^{-3}/\Ef$} for the hard
    disk, up to \mbox{$dt_\mathrm{SD} = 2.5 \times 10^{-3}/\Ef$} for
    the soft disk, up to \mbox{$dt_\mathrm{UTP} = 5.0 \times 10^{-3}/\Ef$}
    for the UTP, and up to \mbox{$dt_\mathrm{TM} = 1.0 \times
      10^{-2}/\Ef$} for the TM pseudopotential. This means that the
maximum timestep for a calculation with the UTP is two and four
times larger than for the soft and hard disk potentials
respectively. Combining this with the reduced variance we therefore accomplish a
total speedup of at least 15 times by using the UTP instead of the hard and 
soft disks.

To summarize, we have demonstrated the importance of using a
pseudopotential with scattering properties that accurately describe
the contact interaction. For weak interactions we observe that a
divergence in the variance in the local energy severely constrains the
accuracy of DMC simulations with soft or hard disk
potentials. At strong interactions these inaccurate potentials
introduce a significant bias into the results, such that we were
unable to attain the $10^{-4}\Ef$ target accuracy in the ground state
energy.  However the UTP delivers
highly accurate results over the full range of interaction strengths
and additionally offers 15 times better computational performance.  We 
therefore recommend the UTP as an accurate and efficacious tool for studying 
the contact interaction in 2D.

\section{Discussion}
\label{Sec:discussion}

We have developed a high-accuracy pseudopotential for the contact
interaction in 2D, building on the work of \reference{Bugnion14}.  We
have demonstrated that our ultratransferable pseudopotential provides accurate
scattering phase shifts, accurate energies for two harmonically
confined particles, and we have demonstrated its advantages in
many-body simulations. The energies obtained with our UTP are over 10 times 
more accurate in the repulsive branch
of the interaction than is afforded by the hard and soft disk
potentials used in recent studies. Moreover, we have
demonstrated that for many-body systems our pseudopotential delivers a
speedup of at least 15 times in diffusion Monte Carlo computations, on top
of the more accurate result.

The performance and ease of construction of the pseudopotential
suggests that it could be widely applicable across first-principles
methods beyond quantum Monte Carlo.  The pseudopotential formalism has
already been used to study the Coulomb \cite{Lloyd-Williams2015} and
dipolar \cite{Whitehead16} interactions. Although in this work we have
focused on using the pseudopotential to accurately capture the
scattering properties of the contact interaction, our formalism allows
the further improvement of modeling of quantum gases by calibrating the
pseudopotentials to more accurately describe
the scattering properties of the underlying Feshbach resonance interaction.
 To next lowest order in scattering wavevector, this
corresponds to including the effective range term essential for describing 
narrow
Feshbach resonances, which may exhibit exotic breached superfluidity
\cite{Liu03,Forbes05}, or other interactions with non-zero effective
ranges, which are applicable in the study of nucleon reactions
\cite{Babenko12}. Rather than a description in terms of the scattering phase
shift, the pseudopotentials could instead be calibrated to other 
scattering properties. For example, they could be calibrated to the 
cross-section for elastic scattering 
measured experimentally via the thermalization rate, or the inelastic loss 
coefficient, to capture the full 
physical interaction between particles \cite{Chin10}.

\acknowledgments {The authors thank Pablo L\'opez R\'ios and
  Pascal Bugnion for useful discussions. The authors acknowledge the
  financial support of the EPSRC [EP/J017639/1]; LMS acknowledges financial 
support
  from the Cambridge European Trust, VSB Fonds, and De Breed Kreiken
  Innovatie Fonds of the Prins Bernhard Cultuurfonds; and GJC
  acknowledges the financial support of the Royal Society and Gonville
  \& Caius College.  Computational facilities were provided by the University 
of Cambridge High Performance Computing Service. Data used for this paper are 
available online \cite{data}.}

\end{document}